\documentclass[two column, superscriptaddress, groupedaddress, amssymb, amsmath, nobibnotes, nofootinbib, aps, showpacs,prb]{revtex4}
\usepackage[dvips]{graphicx}
\usepackage{graphicx}

\begin{document}
\title{Anomalous change of the magnetic moment direction by hole doping in CeRu$_2$Al$_{10}$}

\author{A. Bhattacharyya}
\email{amitava.bhattacharyya@stfc.ac.uk}
\affiliation{ISIS Facility, Rutherford Appleton Laboratory, Chilton, Didcot, Oxon, OX11 0QX, UK} 
\affiliation{Highly Correlated Matter Research Group, Physics Department, University of Johannesburg, Auckland Park 2006, South Africa}
\author{D. Khalyavin}
\email{dmitry.khalyavin@stfc.ac.uk}
\affiliation{ISIS Facility, Rutherford Appleton Laboratory, Chilton, Didcot, Oxon, OX11 0QX, UK} 
\author{D. T. Adroja} 
\email{devashibhai.adroja@stfc.ac.uk}
\affiliation{ISIS Facility, Rutherford Appleton Laboratory, Chilton, Didcot, Oxon, OX11 0QX, UK} 
\affiliation{Highly Correlated Matter Research Group, Physics Department, University of Johannesburg, Auckland Park 2006, South Africa}
\author{A. M. Strydom} 
\affiliation{Highly Correlated Matter Research Group, Physics Department, University of Johannesburg, Auckland Park 2006, South Africa}
\author{A. D . Hillier}
\author{P. Manual}
\affiliation{ISIS Facility, Rutherford Appleton Laboratory, Chilton, Didcot, Oxon, OX11 0QX, UK} 
\author{T. Takabatake}
\affiliation{Department of Quantum Matter, ADSM and IAMR, Hiroshima University, Higashi-Hiroshima 739-8530, Japan}
\author{J.W. Taylor}
\affiliation{ISIS Facility, Rutherford Appleton Laboratory, Chilton, Didcot, Oxon, OX11 0QX, UK} 
\author{C. Ritter}
\affiliation {Institut Laue Langevin, 6 rue Jules Horowitz, 38042 Grenoble, France}

\date{\today} 

\begin{abstract}
We present a detailed investigation of the hole (3\% Re) doping effect on the polycrystalline CeRu$_{2}$Al$_{10}$ sample  by magnetization, heat capacity, resistivity, muon spin rotation ($\mu$SR), and neutron scattering (both elastic and inelastic) measurements. CeRu$_2$Al$_{10}$ is an exceptional cerium compound with an unusually high Neel temperature of 27 K. Here we study the stability of the unusual magnetic order by means of controlled doping, and we uncover further surprising attributes of this phase transition. The heat capacity, resistivity and $\mu$SR measurements reveal an onset of  magnetic ordering below 23 K, while a broad peak at 31 K (i.e. above $T_N$), has been observed in the temperature dependent susceptibility, indicating an opening of a spin gap above $T_N$. Our important finding, from the neutron diffraction, is that the compound orders antiferromagnetically with a propagation vector $\bf k$ = (1 0 0) and the ordered state moment is 0.20(1)$\mu_B$ along the $b-$axis. This is in sharp contrast to the undoped compound, which shows AFM ordering at 27 K with the ordered moment of 0.39(3)$\mu_B$ along the $c-$axis. Similar to CeRu$_2$Al$_{10}$ our inelastic neutron scattering study on the Re doped shows a sharp spin gap-type excitation near 8 meV at 5 K, but with slightly reduced intensity compared to the undoped compound. Further the excitation broadens and shifts to lower energy ($\le$ 4 meV) near 35 K. These results suggest that the low temperature magnetic properties of the hole doped sample is governed by the competition between  the  anisotropic hybridization effect and crystal field anisotropy as observed in hole-doped CeOs$_2$Al$_{10}$.
\end{abstract}

\maketitle
\section{Introduction}
The Kondo insulators (KI) belong to the class of strongly correlated materials forming a group of  either nonmagnetic semiconductors (i.e. FeSb$_2$) with the narrowest energy gap ($\Delta$ in the Kelvin range) or of semimetals (e.g. CeNiSn, Ce$_3$Bi$_4$Pt$_3$, SmB$_6$ and YbB$_{12}$), both with a heavy-fermion metallic state setting in at  elevated temperature $T \ge \Delta$~\cite{ach,jwa,mk,TT, SY, KU, YU, TS}. The gaps inferred from optical, magnetic, transport and thermodynamics properties are almost an order of magnitude smaller than those obtained by band structure calculations, which is due to the presence of strong electronic correlations~\cite{ld,pr}. As a consequence of a strong $c-f$ hybridization and the formation of a Kondo singlet ground state the Kondo insulating ground state is not compatible with magnetic ordering. Very recently the metal$-$insulator transition and a surprisingly high magnetic phase transition in the orthorhombic Ce based cage type materials  CeT$_2$Al$_{10}$ (T = Os and Ru) have attracted considerable attention both experimentally and theoretically~\cite{ams,dta2}. Antiferromagnetic (AFM) ordering of these Ce compounds is found at higher temperatures than in the isostructural Gadolinium compound which rules out that the magnetic order is triggered  by simple Ruderman-Kittel-Kasuya-Yosida (RKKY) type interactions. This suggests that a novel mechanism has to be formulated for the AFM order exists in CeT$_2$Al$_{10}$. 
\par
The recently discovered heavy-fermion compound CeRu$_2$Al$_{10}$ exhibits an unusual magnetic phase transition at 27 K, which is discerned in the temperature dependence of several physical quantities~\cite{ams,tn}. Resistivity, heat capacity, and the thermal conductivity measurements suggest that the phase transition may be accounted for by the formation of a charge-density wave (CDW) or spin-density wave (SDW), which opens an energy gap over a portion of the Fermi surface~\cite{ams,tn, dta2,chl}. Furthermore the spin gap formation in CeRu$_2$Al$_{10}$ has been confirmed through inelastic neutron scattering (gap = 8meV)~\cite{dta2,dta3,ins}, optical study (gap $\sim$ 40 meV)~\cite{opt} and x-ray photoelectron spectroscopy (gap $\sim$ 58 meV)~\cite{xps}.   For CeRu$_2$Al$_{10}$ the magnetic structure reported by Khalyavin {\it et.al. }with a propagation vector of {\bf k} = (1,0,0) involves a collinear antiferromagnetic alignment of the Ce moments along the $c$ axis of the orthorhombic Cmcm space group with a reduced moment of 0.39(3)$\mu_B$~\cite{ddk}. The moment direction is not governed by the single ion crystal field anisotropy, which would align the moment along the easy $a$-axis, having the highest susceptibility~\cite{chit}, but instead by the anisotropic hybridization~\cite{ddk}. This is a very unusual characteristic of the magnetic ordering in CeRu$_2$Al$_{10}$, because in local-moment rare earth magnetism the single-ion model is a standard and well-proven approach in the quantitative treatment of crystal electric fields and magnetocrystalline anisotropy prevailing in these materials. In non-cubic Kondo insulators, the fascinating ground states emerging from extreme anisotropic hybridization is a topic of current interest~\cite{vg}. The lightly hole-doped system CeOs$_{1.96}$Re$_{0.06}$Al$_{10}$ exhibits antiferromagnetic ordering of the Ce moments below $T_N$ = 21 K. The ordered moments are substantially reduced (0.18$\mu_B$) but preserve the anomalous direction along the $c-$axis, indicating the important role of the $c-f$ hybridization in the anisotropic nature of the exchange interactions~\cite{nd1}.
\par
In the present work, we have investigated the marked change in the physical properties of CeRu$_2$Al$_{10}$ induced by a small amount of hole (3\% Re, CeRu$_{1.94}$Re$_{0.06}$Al$_{10}$) doping using magnetization, resistivity, heat capacity, muon spin relaxation, powder neutron diffraction and inelastic neutron scattering measurements. The observation of magnetic Bragg peaks below 23 K in the neutron diffraction data and the independent observation of coherent oscillations due to an internal field in the time dependence of the longitudinal muon spin relaxation spectra confirms the magnetic origin of the transition. Surprisingly, the refinements of the neutron diffraction data show that the best model corresponds to an AFM order of the Ce moment along the $b-$axis of the Cmcm space group with an ordered state moment of 0.20(1)$\mu_B$ at 1.5 K. The inelastic neutron scattering study reveal the presence of a spin gap of 8 meV at 5 K, indicating that aside from the change in ordered moment direction and its magnitude  found here in our analyses, all other attributes of the order parameter are retained in the doped compound CeRu$_{1.94}$Re$_{0.06}$Al$_{10}$ compared to the stoichiometric CeRu$_{2}$Al$_{10}$.~\cite{ddk}

\begin{figure}[htbp]
\centering
\includegraphics[width = 7 cm]{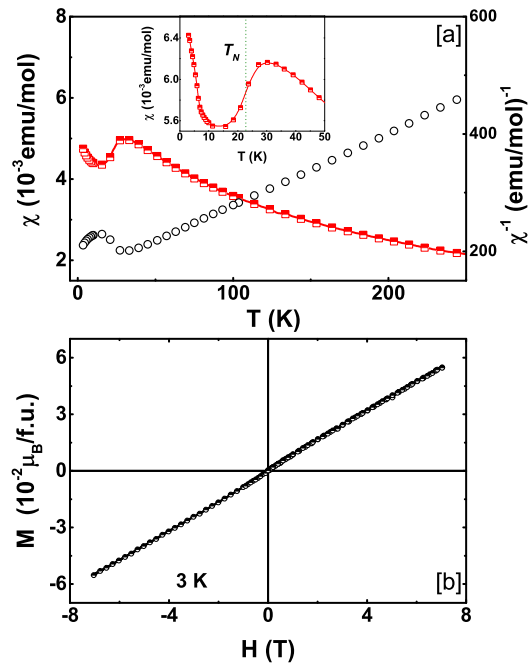}
\caption {(Color online) (a) Temperature dependence of dc magnetic susceptibility ($\chi$ = $M/H$) measured in zero field cooled condition
in the presence of an applied magnetic field of 1 T for CeRu$_{1.94}$Re$_{0.06}$Al$_{10}$. The right scale shows the inverse dc magnetic susceptibility. Inset shows $\chi(T)$ data within temperature range 2 $-$ 50 K. It is clear from $\chi(T)$ data that hybridization gap appears just above $T_N$. (b) The isothermal field dependence of magnetization at 3 K.}
\end{figure}

\begin{figure}[htbp]
\centering
\includegraphics[width = 7 cm]{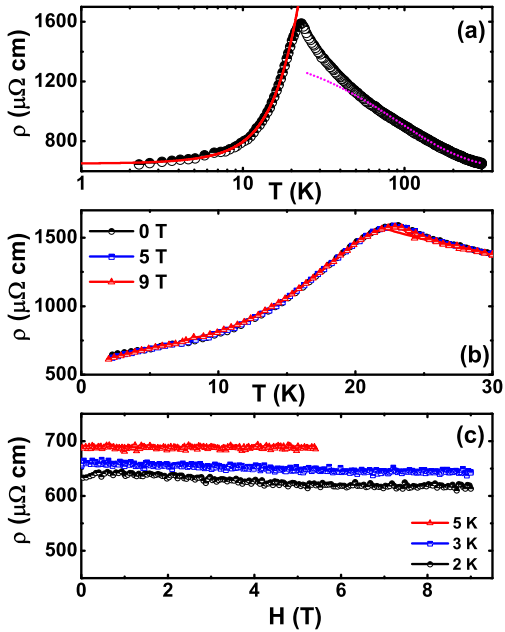}
\caption {(Color online) (a) Semilogarithmic plot of electrical resistivity vs temperature for CeRu$_{1.94}$Re$_{0.06}$Al$_{10}$. The dotted line on the high temperature data illustrates activation type behavior of $\rho(T)$. Below 23 K an antiferromagnetic spin wave expression (see text) represents the data well (red line). (b) and (c) show the temperature (in presence of various applied magnetic field) and field dependence of resistivity for CeRu$_{1.94}$Re$_{0.06}$Al$_{10}$ in the longitudinal configuration.}
\end{figure}

\section{Experimental Details}

Polycrystalline samples were prepared from stoichiometric mixtures of 99.99 \% Ce, 99.99 \% Ru, 99.99 \% Re and 99.999 \% Al by ultra high-purity argon arc melting. The crystal structure was carefully checked and refined using neutron diffraction (see Table I). Magnetic susceptibility measurements were made using a MPMS SQUID magnetometer (Quantum Design). Electrical resistivity, and heat capacity by the relaxation method were performed in a Quantum Design physical properties measurement system (PPMS). The resistivity  in magnetic fields up to 9 T was measured in a longitudinal configuration. Muon spin relaxation ($\mu$SR) and neutron spectroscopy experiments were performed at the ISIS Pulsed Neutron and Muon Facility of the Rutherford Appleton Laboratory, U.K. The $\mu$SR measurements were carried out on the MUSR spectrometer, neutron diffraction  measurements on the WISH diffractometer and inelastic neutron scattering measurements were carried out on the MARI time-of-flight (TOF) spectrometer.~The $\mu SR$ experiments were conducted in longitudinal geometry with a powdered sample and was mounted onto a high purity silver plate. The sample and mount were then inserted into a cryostat with a temperature range of 1.2$-$300 K. In a typical $\mu$SR experiment almost 100\% polarized positive muons are implanted in the sample where, after a short thermalization ($<$ 10$^{-10}$ s), they start precessing about the local magnetic fields. In their decay ($\tau$ = 0.2 $\mu$s) positrons are emitted preferentially in the muon-spin direction at the instant of decay. Muons implanting into any exposed part of the silver mount give rise to a flat time independent background. The asymmetry is calculated by, $G_z(t) =[ {N_F(t) -\alpha N_B(t)}]/[{N_F(t)+\alpha N_B(t)}]$, where $N_B(t)$ and $N_F(t)$ are the numbers of counts at the detectors in the forward and backward positions and $\alpha$ is a constant determined from calibration measurements made in the paramagnetic state with a small (2 mT) applied transverse magnetic field. For the neutron diffraction experiment, a 6 g powder sample was loaded into a cylindrical 6 mm vanadium can and placed in an Oxford Instruments cryostat. Data were recorded in the temperature interval 1.5$-$35 K, with long counting times (8 h) at $T$ = 1.5 K and $T$ = 35 K. Intermediate temperature points were measured with a lower exposition time (2h). The program FULLPROF~\cite{ndr} was used for Rietveld refinements and group theoretical calculations were performed with the aid of the ISOTROPY software~\cite{iso}.

\section{EXPERIMENTAL RESULTS, ANALYSES, AND DISCUSSION}

\subsection{Magnetization, resistivity and heat capacity}

Fig. 1 (a) shows the magnetic susceptibility $\chi$ = $M/H$ versus temperature ($T$) of CeRu$_{1.94}$Re$_{0.06}$Al$_{10}$ as measured at $H$ = 1 T in the $T$ range 2$-$300 K. $\chi (T)$ shows a broad peak at 31 K, which is well above the magnetic ordering observed through neutron diffraction and $\mu$SR  (discussed in sections B and C). This behavior is in contrast with the observed sharp kink  in $\chi(T)$ at 27 K (at $T_N$) of CeRu$_2$Al$_{10}$,  but very similar to the observed broad peak at 45 K (well above the $T_N$ = 28.5 K) in CeOs$_2$Al$_{10}$~\cite{dta2}. These results indicate that the observed broad peak in $\chi(T)$ at 31 K in CeRu$_{1.94}$Re$_{0.06}$Al$_{10}$ conveys the meaning of an opening of a spin gap above $T_N$. The presence of spin gap well above $T_N$ in CeRu$_2$Al$_{10}$ and CeOs$_2$Al$_{10}$ has been confirmed through inelastic neutron scattering and optical measurements~\cite{dta2,km}. The inverse magnetic susceptibility of CeRu$_{1.94}$Re$_{0.06}$Al$_{10}$ and likewise that of CeRu$_2$Al$_{10}$ exhibits Curie-Weiss behavior above 50 K. A linear least-squares fit to the data of CeRu$_{1.94}$Re$_{0.06}$Al$_{10}$ yields an effective magnetic moment $p_{eff}$= 2.51 $\mu_B$, which is very close to free Ce$^{3+}$-ion value (2.54 $\mu_B$), and a negative paramagnetic Curie temperature $\theta_p$= $-$100 K. The value of magnetic moment suggests that the Ce atoms are in their normal Ce$^{3+}$ valence state. Negative value of $\theta_p$ is indicative of a negative exchange constant and/or the presence of the Kondo effect. Fig. 1 (b) shows the $M$ versus $H$ isotherm recorded at 3 K.  $M-H$ data imply that the net magnetization in the ordered state of CeRu$_{1.94}$Re$_{0.06}$Al$_{10}$ is extremely low. It is far from saturation value as that expected from theoretical calculation ~$gJ$= 2.14 $\mu_B$ for Ce$^{3+}$ ions. This is consistent with our neuron diffraction and muon spin relaxation data presented below. The low values of the observed magnetization are expected for an AFM ground state due to the cancellation of magnetization from different magnetic sublattices of Ce ions. There is no clear sign of field induced transition for CeRu$_{1.94}$Re$_{0.06}$Al$_{10}$ upto  $H$ = 7 T. According to the  magnetization which is  monotonous and linear in $H$ up to 7 T, the spontaneous AFM order remains unperturbed at this field value, which was also found in CeRu$_2$Al$_{10}$ up to 13 T ~\cite{ams1}.

\begin{figure}[t]
\centering
\includegraphics[width = 7 cm]{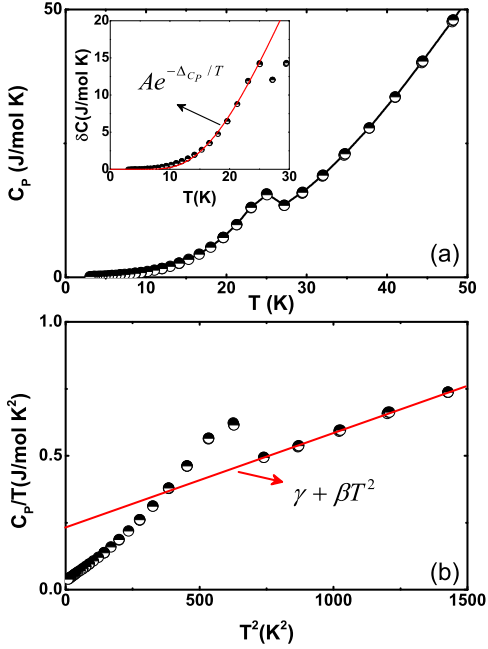}
\caption {(Color online) (a) Temperature variation of specific heat $C_P$ for CeRu$_{1.94}$Re$_{0.06}$Al$_{10}$. Inset (a) displays excess specific heat ($\delta C_P$ vs $T$), $\delta C_P (T)$ = $C_P(T) - \gamma T - \beta T^3$ related to the phase transition as a function of temperature. The solid line is a fit to the data with an expression $\delta C_P (T)$ = A exp($-\Delta _{C_P}/T$). (b) shows a $C_P (T)/T$ vs. $T^2$ plots for CeRu$_{1.94}$Re$_{0.06}$Al$_{10}$.  $C_P (T)$ = $\gamma T$ +$\beta T^3$ is used to fit the experimental data above magnetic ordering.}
\end{figure}

\begin{figure}[htbp]
\centering
\includegraphics[trim = -3cm 0cm 2cm 0cm,width = 8 cm]{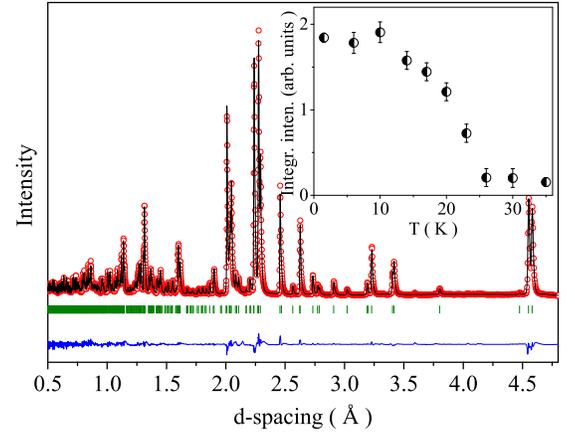}
\caption {(Color online) Rietveld refinement of the neutron powder diffraction pattern of CeRu$_{1.94}$Re$_{0.06}$Al$_{10}$ ($T$=35 K) collected at the backscattering detectors bank (average scattering angle is $154^o$) of the WISH diffractometer. The circle symbols (red) and solid line (black) represent the experimental and calculated intensities, respectively, and the line below (blue) is the difference between them. Tick marks indicate the positions of Bragg peaks in the Cmcm space group. Inset shows the temperature dependence of the integrated intensity of the (1 0 1) magnetic peak.}
\end{figure}

\par

Figs. 2 (a) and (b) show the electrical resistivity $\rho(T)$ of CeRu$_{1.94}$Re$_{0.06}$Al$_{10}$ sample in zero field as well as presence of various applied magnetic field along the electrical current. Data of the resistivity were obtained using a standard four probe method. For CeRu$_{1.94}$Re$_{0.06}$Al$_{10}$ the resistivity below room temperature shows activated type behavior with a narrow gap of 48 K before the onset of a sharp peak at 23 K. By fitting the $\rho(T)$ data above $T_N$ with the formula $\rho$= $\rho_0$ exp($\Delta$/2k$_B T$ ), the value of $\Delta$/k$_B$ is estimated to be 48 K. We interpret the drop in $\rho(T)$ below its 23 K peak in terms of the antiferromagnetic spin-wave gap ($\Delta_{sw}$) expression~\cite{res}
\begin{equation}
\rho(T) = \rho_0+aT\left(1+\frac{2k_BT}{\Delta_{sw}}e^{-\Delta_{sw}/k_BT}\right)+AT^2
\end{equation}
The least-squares fit gives $\Delta_{sw}/k_B$ = 39 K, $\rho(0)$ = 651 Ohm.cm, and $a$ = 54 Ohm.cm/K, for the least squares fit parameters. The Fermi liquid $T^2$ term represents low temperature electron-electron scattering in a metal, but the $A$ coefficient for this term turned out to be negligible in the fit. Fig. 2 (c) shows isothermal $\rho$ versus $H$ data at 2 K, 3 K and 5 K.  The magnetic field dependence of $\rho$ is small at low magnetic fields below $\sim$ 10 T. For CeRu$_2$Al$_{10}$ the magnetoresistance small at low magnetic fields below 20 T at low temperatures, $\rho(H)$ exhibits a negative magnetoresistance at high temperatures~\cite{MR}. A negative magnetoresistance such as this is symptomatic of the conduction electron scattering by the localized magnetic moments.

\par

The temperature dependence of specific heat is shown in Fig. 3 (a). The midpoint (23 K) of the jump in $C_P$ was taken as $T_N$. For CeRu$_2$Al$_{10}$ sample, $C_P/T$ jumps at $T_N$ and the extrapolation of the plot of $C_P/T$ vs $T^2$ to $T$ = 0 gives the Sommerfeld coefficient $\gamma$ of 0.246 J/K$^2$ mol~\cite{chl}. For small amount of hole doping (CeRu$_{1.94}$Re$_{0.06}$Al$_{10}$), $T_N$ decreases to 23 K which agrees with magnetization and $\rho(T)$ data and jumps becomes smaller.  As shown in Fig. 3 (b), the specific heat above 23 K obeys $C_P (T)$ = $\gamma T + \beta T^2$ with $\gamma$ = 0.233 J/K$^2$ mol. The large value of $\gamma$ indicates that CeRu$_{1.94}$Re$_{0.06}$Al$_{10}$ would belong to the family of heavy-fermion systems if the gap was not opened.

\par
The excess specific heat $\delta C_P (T)$ = $C_P(T) - \gamma T - \beta T^3$ can be well described by a thermally activated form $\delta C_P (T)$ = A exp($-\Delta_{C_P}/T$) with A = 200 J /mol K and $\Delta_{C_P}$ = 75 K = 3.26 $T_N$ (solid line in the upper inset of Fig. 3 (a)). For CeRu$_2$Al$_{10}$ the relation between energy gap and $T_N$ is $\Delta C_P$ = 100 K = 3.74 $T_N$~\cite{chl}. The specific-heat data clearly reveal that the anomaly associated with the second-order phase transition ($\lambda$ type anomaly) involves an energy gap of about 75 K. The origin of the energy gap could be either a gap over the part of the Fermi surface or anisotropic gap in the spin wave due to combination of single ion anisotropy and anisotropic exchange. It is noted that the extracted energy gap is also comparable to the value of 8 meV peak from the  inelastic neutron-scattering experiment. The above experimental results suggest that both CeRu$_2$Al$_{10}$ and CeRu$_{1.94}$Re$_{0.06}$Al$_{10}$ show similar phase transitions with an opening of a spin gap below  $T_N$.

\begin{table}[b]
\begin{center}
\caption{Structural parameters of CeRu$_{1.94}$Re$_{0.06}$Al$_{10}$ refined from the neutron diffraction data collected at $T$ = 35 K in the orthorhombic Cmcm  [$R_{bragg}$ = 3.98\%) space group. Occupancies for all the atoms in the refinement procedure were fixed to the nominal chemical content.}

\begin{tabular}{lccccccccccccc}
\hline
\hline
Atom && Site &&  x  && y  && z &&  &&\\ 
\hline
Ce 	&& 	4c  && 0 && 0.1265(3)  && 0.25 &&  && \\ 
Ru/Re && 8d  && 0.25 && 0.25  && 0 && && \\ 
Al1 && 	8g  && 0.2237(3) && 0.3625(1) && 0.25 &&  && \\ 
Al2 && 	8g && 0.3490(2) && 0.1314(3)&& 0.25 &&  && \\ 
Al3 && 	8f && 0 && 0.1616(4)  && 0.5986(4) && && \\ 
Al4 && 	8f && 0 && 0.3796(3)  && 0.0496(1) &&  && \\ 
Al5 && 	8e && 0.2291(3) && 0  && 0 && && \\ 
\hline 
\hline
 Ce(Ru$_{1-x}$&&  a  && b  && c&& V\\ 
Re$_x$)$_2$Al$_{10}$ &&  (\AA) && (\AA)  && (\AA) && (\AA$^3$)\\ 
\hline
$x$=0  &&  9.1322(2) && 10.2906(1)  && 9.1948(2)&& 864.08  \\
$x$=0.03   && 9.1292(2) && 10.2870(1)  && 9.1914(2) && 863.185 \\ 
\hline 
\hline
\end{tabular}
\end{center}
\end{table}

\subsection{Neutron diffraction: onset of magnetic long-range order}

The neutron-diffraction patterns collected  above 23 K (as shown in Fig. 4) for CeRu$_{1.94}$Re$_{0.06}$Al$_{10}$ are consistent with the Cmcm symmetry of the nuclear structure and can be satisfactorily fitted with the structural model proposed by Thiede {\it et al.}~\cite{ddk,vmt,nd}. The structural parameters are listed in Table I. Below 23 K, a set of new weak reflections associated with the propagation vector {\bf k}= (1,0,0) (Y point of symmetry in Miller and Love notations~\cite{scm}) appears indicating the phase transition detected by $\chi(T)$, $\rho(T)$, $C_P(T)$ and $\mu$SR techniques. The origin of the transition most probably is due to a long range magnetic ordering of the Ce moments as observed in the undoped CeRu$_2$Al$_{10}$ compound. This conclusion is supported by the fact that the additional reflections are only clearly visible at low-momentum transfer, indicating that they follow the squares of the magnetic form factor $F^2(Q)$ (here it is Ce$^{3+}$ form factor). Another important feature is the reduction in the background observed at low $Q$ and concomitant with the appearance of the magnetic peaks below 23 K. Once again, this is consistent with the existence of a magnetic transition below which the intensity contained in the paramagnetic scattering is suppressed and transferred to magnetic Bragg scattering. The temperature dependence of the integrated intensity of the strongest (101) magnetic Bragg peak is shown in Fig. 4 (inset).

\begin{figure}[t]
\centering
\includegraphics[width = 7 cm]{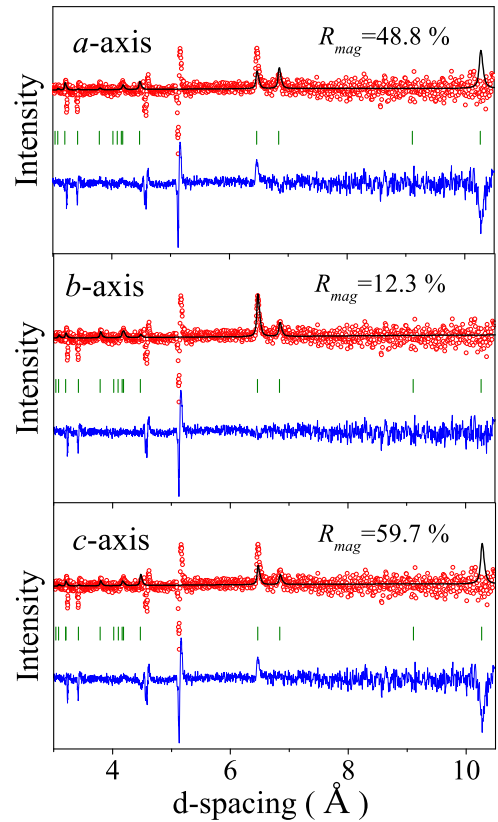}
\caption {(Color online) Rietveld refinements of the magnetic intensity of the CeRu$_{1.94}$Re$_{0.06}$Al$_{10}$ composition obtained as a difference between the diffraction patterns collected at 1.5 K and 35  K. The circle symbols (red) and solid line represent the experimental and calculated intensities, respectively, and the line below (blue) is the difference between them. Tick marks indicate the positions of Bragg peaks for the magnetic scattering with the (${\bf k}=1,0,0$) propagation vector. The refinement quality is demonstrated for three models, with moments along the $a$-axis (top), $b$-axis (middle) and $c$-axis (bottom). The asymmetric features are due to slight differences in $d$ spacing of the nuclear peaks at different temperatures (thermal expansion).}
\end{figure}

\begin{figure}[t]
\centering
\includegraphics[width = 7 cm]{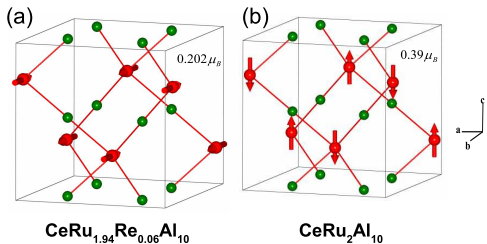}
\caption {(Color online) Magnetic structures of the hole-doped (found in this paper) and CeRu$_2$Al$_{10}$~\cite{ddk} samples. For clarity, only Ce and Ru/Re atoms are shown (top).}
\end{figure}

\par

By  taking  the  difference between the diffraction patterns collected  at  $T$ = 1.5  K  and  $T$ = 35 K, several magnetic Bragg peaks with intensities significantly higher than the error bars  [Fig. 5] can be identified. The important observation from the difference plot is the absence of the magnetic (010) peak near 10.3 \AA~, which is clearly seen in pure CeRu$_2$Al$_{10}$, Ce(Ru$_{1-x}$Fe$_x$)$_2$Al$_{10}$, $x$ = 0.4 and CeOs$_2$Al$_{10}$~\cite{dta3,DTA1}. Taking into account that neutron scattering intensity is proportional to  square of the moment component perpendicular to the scattering vector, this observation conforms that the magnetic moment direction in CeRu$_{1.94}$Re$_{0.06}$Al$_{10}$ is along the $b-$axis and not along $c-$axis as observed in CeRu$_2$Al$_{10}$, CeOs$_2$Al$_{10}$ and CeOs$_{1.94}$Re$_{0.06}$Al$_{10}$ ~\cite{nd1}. Further it is to be noted that the moment direction in Ce(Os$_{1-x}$Ir$_x$)$_2$Al$_{10}$, $x$=0.08 is along the $a-$axis (moment value of 0.92(1)$\mu_B$)~\cite{ndd}. In spite of the different moment direction, the observed magnetic Bragg peaks in CeRu$_{1.94}$Re$_{0.06}$Al$_{10}$ can be indexed based on the propagation vector {\bf k}=(1 0 0), which is the same as observed in pure CeRu$_2$Al$_{10}$~\cite{ddk}.

\par
To obtain an appropriate model for the magnetic structure of CeRu$_{1.94}$Re$_{0.06}$Al$_{10}$, we employed a method whereby combinations of axial vectors localized on the 4$c$(Ce) site and transforming as basis functions of the irreducible representations of the wave-vector group [{\bf k}=(1,0,0)], are systematically tested~\cite{nds}. The symmetry analysis reveals that the reducible magnetic representation is decomposed into six one-dimensional representations, labeled $Y^+_i$ ($i$ = 2, 3, 4) and $Y^-_i$ ($i$ = 1, 2, 3).  The $Y^+_i$ representations imply FM alignment of the Ce moments within the primitive unit cell, along different crystallographic directions. On the other hand, $Y^-_i$ transform Ce moments which are AFM coupled within the primitive unit cell. In agreement with the Landau theory of continuous transitions, we found that a single irrep is involved. A unique solution associated with the $Y^-_1$ representation was found to provide an excellent refinement quality of the magnetic intensity (Fig. 5 (middle)). The refinement yields the ordered state of Ce moments along the $b$-axis to be 0.20(1)$\mu_B$ (see Fig. 6 (a), where the magnetic unit cell is shown). On the other hand the fit to the data with the moments along either the $a-$axis [Fig. 5 (top)] or the $c$-axis [Fig. 5 (bottom)] results in a much worth refinement quality. Thus, there are two important differences to be noted compared with parent CeRu$_2$Al$_{10}$, (1) the ordered moments value is almost half in the present sample compare to undoped one and (2) the direction of the moments is along the $b-$axis and not along the $c-$axis as in undoped CeRu$_2$Al$_{10}$ and also in CeOs$_{1.94}$Re$_{0.06}$Al$_{10}$~\cite{ddk,nd1} see Fig. 6(b).

\par

\begin{figure}[b]
\centering
\includegraphics[width = 8 cm]{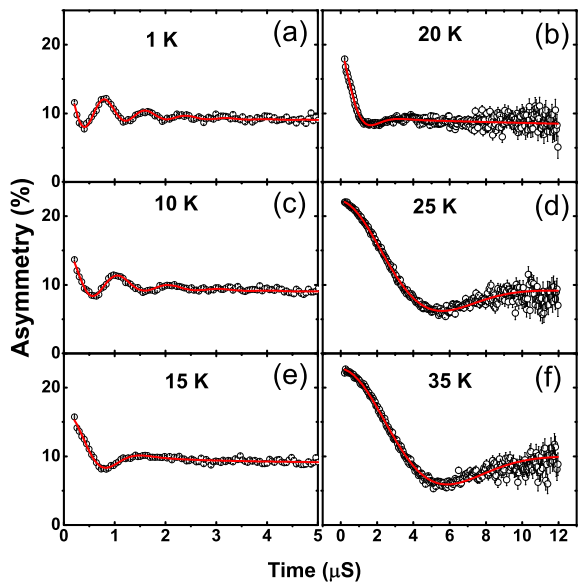}
\caption {(Color online) The time evolution of the muon spin relaxation for various temperatures (above and below $T_N$) in zero field. The solid line is a least-squares fit [using Eq (2) (above $T_N$) and Eq (3) (below $T_N$)] to the data as described in the text.}
\end{figure}

\subsection{Muon spin relaxation: evidence of small magnetic moment in the ordered state}

To investigate the microscopic magnetic nature at $T_N$ we have measured positive muon-spin relaxation on CeRu$_{1.94}$Re$_{0.06}$Al$_{10}$,  using polycrystalline samples in the temperature range between 1.5 and 50 K. The $\mu$SR technique is sensitive to local magnetic order via the decrease in muon asymmetry spectra and the enhanced muon-spin relaxation at the transition temperature. For CeRu$_2$Al$_{10}$ two muon frequencies (10.5 mT and 3 mT) were observed, suggesting that muons occupy at least two different sites in the sample~\cite{ddk}. One of these components also exhibits unusual behaviors, showing a dip at about 11 K in the temperature dependence, and was suggested to be probably related to a structure distortion or resistivity anomaly. For CeRu$_{1.94}$Re$_{0.06}$Al$_{10}$ we found Kubo-Toyabe type behavior above $T_N$ (like CeRu$_2$Al$_{10}$) and below $T_N$ its shows clear signature of time oscillation consistent with a very small (0.20(1)$\mu_B$).

\par
Detailed $\mu$SR data were obtained in zero external field in order to study the variation of $B_{int}$ with $T$. Figs. 7 (a)-(f) show the zero-field time dependence asymmetry spectra at various temperatures below and above magnetic ordering of CeRu$_{1.94}$Re$_{0.06}$Al$_{10}$. It is clear from Fig. 7 that small amount of hole doping drastically changes the time dependence of asymmetry spectra compared with the pure compound. Above 23 K, we observe Kubo-Toyabe (KT) type behavior ~\cite{musr1}i.e. a strong damping at shorter time, and the recovery at longer times, arising from a static distribution of the nuclear dipole moment. On the time scale of the muon, these nuclear spins are static and randomly orientated. The spectra above the phase transition temperature are best described by Kubo-Toyabe times exponential decay signals plus background as shown in Figs. 7 (d) and (f): 

\begin{equation}
G_{z_1}^{KT}(t) =\frac{A_2}{3}[1+2(1-\sigma_{KT}^2t^2)e^{(-\frac{\sigma_{KT}^2t^2}{2})}]e^{-\lambda_2 t}+A_{bg}
\end{equation}

where initial amplitude of the KT decay is $A_2$;  $\lambda_2$ is the relaxation rate associated with the dynamic electronic spin fluctuations; $A_{bg}$ is a constant background arising from muon stopping on the silver sample holder. $A_{bg}$ was estimated from 50 K data and kept fixed for fitting all the other spectra. Nuclear depolarization rate is $\sigma_{KT}$, and $\sigma_{KT}/\gamma_{\mu}$ = $\Delta$ is the local Gaussian field distribution width, $\gamma_{\mu}$ is the gyromagnetic ratio of the muon. $\sigma_{KT}$ was found to be almost temperature independent as shown in Fig. 8 (d) with its value equal to 0.29 $\mu S^{-1}$. Using a similar $\sigma_{KT}$ value  Kambe {\it et. al.}~\cite{S. Kambe} have suggested $4a$ (0,0,0) as the muon stopping site in CeRu$_2$Al$_{10}$, while for CeOs$_2$Al$_{10}$~\cite{DTA1,ab}, the muon stopping site was assigned to the $4c$ (0.5, 0, 0.25) position. Dipolar fields calculation by Guo {\it et. al.}on Ce(Ru$_{1-x}$Rh$_x$)$_2$Al$_{10}$ ($x$ = 0$-$0.08) ~\cite{musr2} suggest two stopping sites $4c$ and $4a$, in the latter the the internal field is almost zero. Our dipolar fields calculation for CeRu$_{1.94}$Re$_{0.06}$Al$_{10}$ suggest a muon stopping site is $4c$ (0.5, 0, 0.25).

As the temperature is approached near magnetic transition, the $\mu$SR spectra clearly show the presence of coherent oscillations. Below 23 K, all spectra are well described uniformly by the phenomenological function
\begin{equation}
G_{z_2}(t) = A_1cos(\omega t+\phi)e^{-\lambda_1 t}+ G_{z_1}^{KT}(t)
\end{equation}

where $\sigma_{KT}$ = 0, $G_{z_1}^{KT}(t)$ term becomes  $A_2$$e^{-\lambda_2 t}$+$A_{bg}$, $\lambda_1$ is the muon depolarization rate (arising from the distribution of the internal field), $\phi$ is the phase and $\omega$ = $\gamma_\mu B_{int}$ is the muon precession frequency ($B_{int}$ is the internal field at the muon site). The first term represents the transverse components of the internal fields seen by the muons along which they precess, while the second term represents the longitudinal component. 

\par
The temperature dependencies of these parameters are shown in Figs. 8 (a)-(d). Below 23 K, as shown in Fig. 8 (a) there is a loss of initial asymmetry in $A_2$ compared to that of the high temperature value. The initial asymmetry associated with frequency term $A_1$ starts to increase below this temperature (23 K) [see Fig. 8 (a)], indicating the onset of a long-range ordered state in CeRu$_{1.94}$Re$_{0.06}$Al$_{10}$ which agrees with the specific heat and magnetic susceptibility data. The value of the of  Lorentzian decay term $\lambda_2$ starts to increase below 23 K. $\lambda_1$ associated with oscillating term on the other hand remains almost constant. Fig. 8 (d) shows the temperature dependence of the muon depolarization rate which seems temperature independent. Fig. 8 (b) shows the temperature dependence of the internal field (or muon precession frequency) at the muon site. This shows that the internal fields appear below 23 K, signifying clear evidence for long-range magnetic order. However, the associated internal fields are found to be same order of magnitude compared to CeRu$_2$Al$_{10}$ ($B_{int}$ = 10 mT at base temperature). In order to find out the nature of the magnetic interaction temperature dependence of internal field was fitted~\cite{ab1}

\begin{equation}
B_{int}(T)= B_0\left[1-\left(\frac{T}{T_N}\right)^{\alpha}\right]^{\beta}
\end{equation}

Observed parameters are $\beta$ = 0.96(2),  $B_0$ = 9.4(3) mT, $\alpha$ = 1.47(2) and $T_{N}$ = 23 (3) K.  A fit with $\beta \sim$ 0.96(2) suggests the magnetic interactions in hole doped system is non-mean field like behavior. $\alpha >$  1 indicates complex magnetic interactions in this system~\cite{dikh,sb}.

\begin{figure}[t]
\centering
\includegraphics[width = 9 cm]{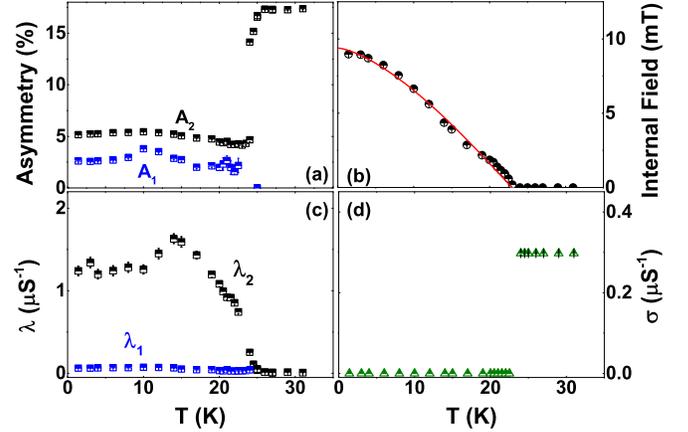}
\caption {(Color online) The temperature dependence of (a) the initial asymmetries $A_1$ and $A_2$  (b) the internal field at the muon site  (c) the depolarization rates $\lambda_1$ and $\lambda_2$  (d) the depolarization rate $\sigma_KT$. The solid line in (c) is fit to the data using Eq (4)(see text).}
\end{figure}

\begin{figure}[t]
\centering
\includegraphics[width = 7 cm]{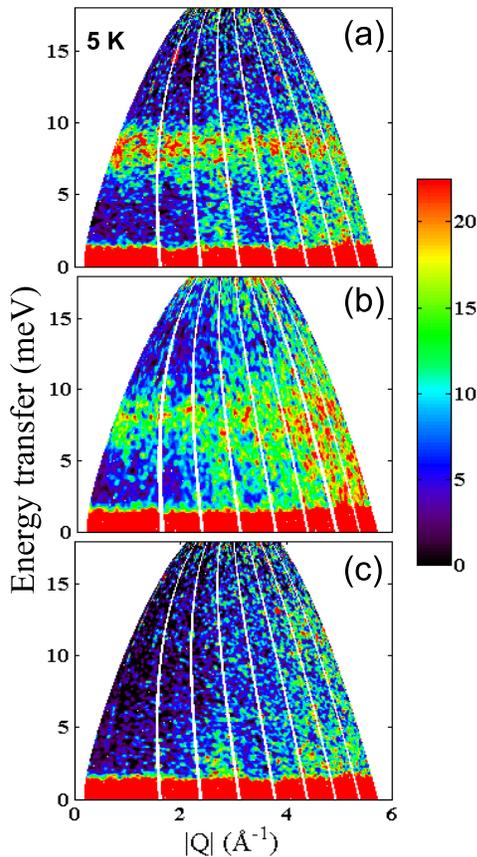}
\caption {(Color online) Color-coded inelastic neutron scattering intensity of (a) CeRu$_2$Al$_{10}$ (b) CeRu$_{1.94}$Re$_{0.06}$Al$_{10}$ and (c) non-magnetic LaRu$_2$Al$_{10}$, at 5 K, measured with incident energy of $E_i$ = 20 meV on the MARI spectrometer.}
\end{figure}

\subsection{Inelastic neutron scattering study}

INS measurements on CeRu$_{2}$Al$_{10}$ clearly revealed the presence of a sharp inelastic magnetic excitation near 8 meV below 29 K, due to opening of a gap in the spin-excitation spectrum, which transforms into a broad response at and above 35 K ~\cite{DTA1,nd}.  Hole doping results dramatic change in magnetic moment directions  and its value is reduced to almost half compared to undoped system. It is of great interest to study inelastic neutron scattering to see how the spin gap and its $Q$ and temperature dependencies vary for CeRu$_{1.94}$Re$_{0.06}$Al$_{10}$. In this section we briefly report the temperature dependence of low energy INS spectra of CeRu$_{1.94}$Re$_{0.06}$Al$_{10}$. We have also measured the nonmagnetic phonon reference compound LaRu$_2$Al$_{10}$.

\par
Figs. 9 (a)$-$(c) show the color plots of the total scattering intensity (magnetic and phonon contributions), energy transfer {\it vs.} momentum transfer measured at 5 K for CeRu$_{2}$Al$_{10}$, CeRu$_{1.94}$Re$_{0.06}$Al$_{10}$ and LaRu$_2$Al$_{10}$. For  CeRu$_{1.94}$Re$_{0.06}$Al$_{10}$  compound we find clear magnetic excitation or spin gap energy around 8 meV. This value is in agreement with the spin gap energy estimated from the specific heat and resistivity studies as discussed earlier. Further for comparison we have also shown the scattering from the nonmagnetic reference compound LaRu$_2$Al$_{10}$ which confirms the magnetic nature of the excitations in the Ce compounds.  We have plotted the data in one-dimensional (1D) ($Q$-integrated between 0 and 3 \AA) energy cuts (see Figs. 10 (a)-(c)) taken from the two-dimensional (2D) plots. It is clear from this 1D cuts that the position of the spin gap excitation remains nearly the same in both the compounds while the linewidth of the excitation increases in the Re-doped systems: $\Gamma$ = 1.1 meV for CeRu$_2$Al$_{10}$ and 1.6 meV for CeRu$_{1.94}$Re$_{0.06}$Al$_{10}$. The estimated value of the susceptibility ($\chi_{INS}$) at 5 K is 3.6$\times$10$^{-3}$ emu/mol for CeRu$_{1.94}$Re$_{0.06}$Al$_{10}$ and 5.1$\times$10$^{-3}$ emu/mol for CeRu$_{2}$Al$_{10}$.  At 25 K and 35 K the excitation becomes broad, but still keep inelastic nature.

\par
Now we compare the effect of reduction of the moment and its direction on the stability of the spin gap type excitations in the doped CeRu$_2$Al$_{10}$ and CeOs$_{2}$Al$_{10}$.  First we would like to mention that in the present case for both doped (CeRu$_{1.94}$Re$_{0.06}$Al$_{10}$) and undoped (CeRu$_2$Al$_{10}$) systems the moment direction does not align along the CEF anisotropy, which expects moment along $a-$axis, and further the observed moment values are reduced considerably from the free-ion Hund's rule values, 0.20(1)$\mu_B$ and 0.39 (3)$\mu_B$, respectively. Results of these, we have seen well defined spin gap excitations in both the system having small moment that does not follow CEF anisotropy.  Our preliminary measurements of inelastic neutron scattering on electron doped (i.e. 10\% Rh) in CeRu$_2$Al$_{10}$ reveals that spin gap value is reduced to 5 meV with considerable reduction in the intensity with broaden linewdith, which make it difficult to detect the spin gap clearly~\cite{dta4}. The study of $\mu$SR on Ce(Ru$_{1-x}$Rh$_x$)$_2$Al$_{10}$ clearly indicates that the internal fields are higher in the Rh-doped samples ~\cite{musr2}, which directly implies larger moment value. A very similar increase in the moment value, 1$\mu_B$, with direction of moment along $a-$axis has been observed in Ce(Os$_{1-x}$Ir$_x$)$_2$Al$_{10}$ and inelastic study in this case does not reveal any sign of spin gap down to 2 K and also down to 0.6 meV~\cite{ab}. These observations indicate that when the ordered state moment value is larger that destabilizes the spin gap formation. This implies that when the ordered moment value is small, and moreover not aligned with the CEF easy axis, then the $c-f$ hybridization is active and stabilizes the spin gap, while when moment value is larger (and follows CEF anisotropy) the spin gap becomes unstable. 

\begin{figure}[t]
\centering
\includegraphics[width = 6 cm]{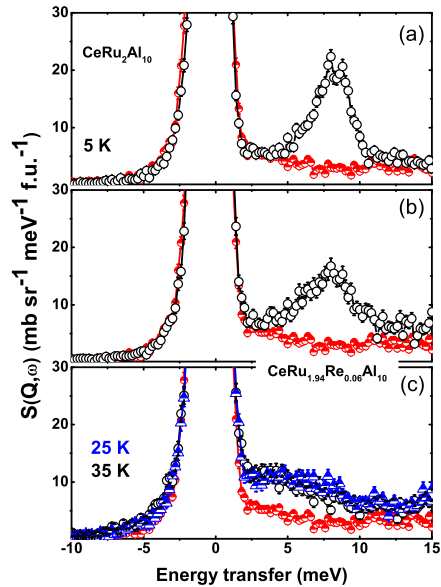}
\caption {(Color online) $Q$-integrated (0$\le$$Q$$\le$2.5~\AA) intensity versus energy transfer of (a) CeRu$_2$Al$_{10}$ (b) and (c) CeRu$_{1.94}$Re$_{0.06}$Al$_{10}$ along with nonmagnetic phonon reference compound LaRu$_2$Al$_{10}$ (red half filled circles), measured with respective incident energy of $E_i$ = 20 meV.}
\end{figure}

\subsection{Concluding Remarks}

Our results provide compelling evidence that the CeRu$_{1.94}$Re$_{0.06}$Al$_{10}$ exhibits a phase transition at 23 K associated with long-range magnetic ordering of the Ce sublattice. The propagation vector of the ordered state is $\bf k$ = (1,0,0) and does not change with temperature. The magnetic structure at $T$ = 1.5 K involves a collinear antiferromagnetic orientation of the Ce moments along the $b-$axis of the Cmcm space group with a magnitude of 0.20(1)$\mu_B$. 

\par

Further inelastic neutron scattering study reveals a clear sign of spin gap type excitation with energy scale of 8 meV in the hole doped system, which is same as that observed in the undoped CeRu$_2$Al$_{10}$. Our present hole doping study  along with other  investigations on the doped systems of CeT$_2$Al$_{10}$ (T = Ru and Os) indicate that the smaller value of the ordered state moment (also direction not governed by CEF anisotropy) stabilizes the spin gap formation. On the other hand the larger value ($\sim$1 $\mu_B$) of the ordered state moment (governed by CEF anisotropy) destabilizes the spin gap ground state. It has been reported through magnetization study that the ordered state moment is along the $b-$axis in Ce$_{1-x}$La$_x$Ru$_2$Al$_{10}$  ($x$ =0.1) at ambient pressure. This has been attributed to a negative chemical pressure effect~\cite{ht}. However, in the present hole-doped study the volume contracts by 0.1\%, concomitant with a positive chemical pressure. Considering these two observations we propose that change in the moment direction in the hole doped CeRu$_2$Al$_{10}$ is due to electronic effect. It is expected that our findings in this work will generate new theoretical interest that might help to understand the real mechanism of spin gap formation and higher magnetic ordering temperature of this family of compounds.   

\subsection*{}
We would like to thank  Prof. P. Riseborough, Dr. Jean-Michel Mignot and Prof. Y. Muro for an interesting discussion. A.B would like to acknowledge FRC of UJ, NRF of South Africa and ISIS-STFC for funding support. D.T.A., and A.D.H. thanks the CMPC-STFC, grant number CMPC-09108, for financial support.  A.M.S. thanks the SA-NRF (Grant 78832) and UJ Research Committee for financial support. T.T. thanks KAKENHI No.26400363 from MEXT, Japan.

\end{document}